\documentclass[aps,prc,twocolumn,superscriptaddress,showpacs]{revtex4}
\usepackage{bm}

\newcommand{\vct}[1]{{\bf #1}}
\newcommand{\gvct}[1]{\bm{#1}}
\newcommand{\fslash}[1]{\!\not\!#1}
\newcommand{\kf}{k_{\rm F}}
\newcommand{\ekf}{E_{\rm F}}
\newcommand{\kfp}{k_{\rm p}}
\newcommand{\kfn}{k_{\rm n}}
\newcommand{\kfi}{k_{\rm i}}
\newcommand{\vf}{v_{\rm F}}
\newcommand{\mitg}{{\mit\Gamma}}
\newcommand{\thp}[1]{\theta^{\rm (p)}_{\vct{#1}}}
\newcommand{\thn}[1]{\theta^{\rm (n)}_{\vct{#1}}}
\newcommand{\thi}[1]{\theta^{\rm (i)}_{\vct{#1}}}
\newcommand{\ep}[1]{E_{\vct{#1}}}

\begin{document}
\title{
Role of the Landau-Migdal Parameters with
the Pseudovector
and
the Tensor Coupling
in Relativistic Nuclear Models\\
-- The Quenching of the Gamow-Teller Strength --
}

\author{Tomoyuki Maruyama}
\affiliation{
College of Bioresource Sciences,
Nihon University,
Fujisawa 252-8510, Japan}
\affiliation{
Advanced Science Research Center,
Japan Atomic Energy Research Institute,
Tokai 319-1195, Japan
}

\author{Haruki Kurasawa}
\affiliation{
Department of Physics, Faculty of Science, Chiba University, Chiba
263-8522, Japan
}

\author{Toshio Suzuki}
\affiliation{
Department of Applied Physics, Fukui University, Fukui 910-8507, Japan
}

\begin{abstract}
Role of the 
Landau-Migdal parameters with the pseudovector ($g_a$)
and the tensor coupling ($g_t$) is examined for the giant
Gamow-Teller (GT) states in the relativistic random phase
approximation (RPA). The excitation energy is dominated by both
$g_a$ and $g_t$ in a similar way, while the GT strength 
is independent of $g_a$ and $g_t$ in the RPA of the nucleon space, 
and is quenched, compared with that in non-relativistic one.
The coupling of the particle-hole states
with nucleon-antinucleon states is 
expected to quench the GT strength further through $g_a$.
\end{abstract}

\pacs{21.60.Jz, 21.65.+f, 24.30.Gd}
%% \keywords{Gamow-Teller states}

\maketitle
 
\section{Introduction}\label{intro}

The long- and medium-range parts of the nuclear spin-isospin interaction
are well described by the one-meson exchange approximation.
In the non-relativistic limit,
they are written in the momentum space as\cite{brown}
\begin{eqnarray}
V_{\pi+\rho}&=&
-\left[\left(\frac{f_\pi}{m_\pi}\right)^2 
 \frac{(\gvct{\sigma}_1\cdot\vct{q})(\gvct{\sigma}_2\cdot\vct{q})}{\vct{q}^2 
 +m_\pi^2} \right. \nonumber \\
 & &
  \left.
+\left(\frac{f_\rho}{m_\rho}\right)^2 
 \frac{(\gvct{\sigma}_1\times\vct{q})\cdot
 (\gvct{\sigma}_2\times\vct{q})}{\vct{q}^2 
 +m_\rho^2}\right]\gvct{\tau}_1\cdot\gvct{\tau}_2,\label{int}
\end{eqnarray}
where the first term stands for the one-pion exchange potential,
while the second term the one-rho exchange potential.
They are obtained from the pseudovector and tensor meson-nucleon
couplings, respectively. In the short-range part of the interaction,
on the other hand,
many-body correlations should be taken into account.
Their effects
are approximately expressed by the Landau-Migdal (LM) parameter
$g'$\,\cite{brown},
\begin{eqnarray} 
V_{{\rm LM}}
 =\left(\frac{f_\pi}{m_\pi}\right)^2g'\gvct{\sigma}_1\cdot\gvct{\sigma}_2\, 
\gvct{\tau}_1\cdot\gvct{\tau}_2.\label{LM} 
\end{eqnarray}
Experimentally the value of $g'$ is estimated to be
about 0.6\cite{ss}.
This fact is understood in such a way that  about a half of the value
cancels the zero-range part of the one-pion exchange potential, while
the rest the short-range part of the rho-meson
exchange potential in Eq.(\ref{int})\cite{shimizu,arima}.
The contribution from the exchange terms of the one-pion and one-rho
potential is also considered to be included in $g'$.
Indeed, if we calculate the LM parameter so as to cancel
the zero-range part of the one-pion potential and to include the
contribution from the exchange term of the finite-range part,
we obtain from Eq.(\ref{int})
\begin{eqnarray}
 g'_\pi=\frac{1}{3}\left(1+\frac{m_\pi^2}{16k_{\rm F}^2}\ln
		    \frac{m_\pi^2}{m_\pi^2+4k_{\rm F}^2}\right).
\end{eqnarray}
This value is about 0.318 for the pion mass $m_\pi=140$ MeV and the
 Fermi momentum $k_{\rm F}=1.36$fm$^{-1}$.  
The corresponding expression from the rho-meson exchange potential
is given by
\begin{eqnarray}
 g'_\rho=\frac{2}{3}\left(\frac{f_\rho}{m_\rho}\right)^2
  \left(\frac{m_\pi}{f_\pi}\right)^2
  \left(1+\frac{m_\rho^2}{16k_{\rm F}^2}\ln
		    \frac{m_\rho^2}{m_\rho^2+4k_{\rm F}^2}\right).
\end{eqnarray}
If we simply calculate this value by using the strong coupling
$f_\rho^2/4\pi=4.86$ together with $m_\rho=770$ MeV, we obtain
$g'_\rho=1.07$. Refs.\cite{shimizu} and  \cite{arima}
have shown, however, that in the $\rho$-meson
exchange, the short-range part of the finite range interaction
should be also taken out,
since the $\rho$-mass is
much larger than the $\pi$-mass. This contribution makes
the above value of $g'_\rho$ smaller as 0.281,
since the finite range part has an opposite sign to that of the
zero-range part.
Thus the experimental value about 0.6
is well understood by the sum of $g'_\pi+g'_\rho$ in
non-relativistic models.   
  
Now, 
for the last 30 years it has been shown that nuclear structure
is well explained phenomenologically by relativistic
models\cite{serot}.
Not only the ground state, but also the excited states are
well reproduced\cite{ring}.  
So far, however, spin-dependent excitations of nuclei have
not been studied in details\cite{conti}.
One of the reasons may be because
there is no estimation of the LM parameter yet
in relativistic models.
Moreover, there is no unique way to extend the above
non-relativistic $g'$ to the relativistic one.

The purpose of the present paper is to introduce phenomenologically
the LM parameter in relativistic models, and to explore
model-dependence of the excitation energy and strength of the giant
Gamow-Teller (GT) resonance state.
We will construct relativistic LM parameters
so as to reproduce $g'$ of Eq.(\ref{LM})
in the non-relativistic limit.
Such candidates will be given by
contact terms with the pseudovector ($g_a$)
and the tensor ($g_t$) coupling in the nuclear Lagrangian.

In the previous paper\cite{ksg1,ksg2, ks1},
we investigated already the LM parameter of the
pseudovector type for the description of
the giant GT resonance states. 
It was shown that the excitation energy 
in non-relativistic models is reproduced and that the GT excitation
strength in the nucleon space is 
quenched, compared with the sum rule value
by Ikeda-Fujii-Fujita (IFF)\cite{ikeda}.
In particular, the quenching, which is
consistent with recent experiment\cite{sakai},
was an important conclusion,
since in conventional non-relativistic models
the value of the GT strength 
is fixed by the IFF sum rule.   
In this paper we will pay our attention mainly
to whether or not those conclusions in the relativistic model
are changed by the LM parameter with the tensor coupling. 

In the following section, we will calculate the correlation function in
the random phase approximation (RPA), where the correlations are
induced through the LM parameters $g_a$ and $g_t$.
In Sec.\ref{energy} and \ref{strength},
the excitation energy and strength of the GT state
will be estimated, respectively. Since
all calculations are performed for nuclear matter, discussions
will be given according to analytic formulae of the
excitation energy and strength. They make clear
structure of the giant GT states in relativistic models.
The final section will be devoted to a
brief conclusion of the present paper.    

\section{Relativistic RPA Correlation Function}

We assume that nucleons and antinucleons
in nuclear matter are bounded in the mean field of the Lorentz
scalar and vector potentials.
The Lorentz scalar potential
will be included into the nucleon effective mass $M^\ast$ below.
On the other hand, 
we will not describe explicitly the Lorentz vector potential
throughout the present paper,
since it does not play any role in description of the GT states.

The particle-hole and nucleon-antinucleon correlations are assumed
to be caused by the Lagrangian, 
\begin{eqnarray}
 {\cal L}
= 
 \frac{g_a}{2}\,
 \overline{\psi}\mitg^\mu_i\psi\,
 \overline{\psi}\mitg_{\mu i}\psi
 +
 \frac{g_t}{4}\,
 \overline{\psi}T^{\mu\nu}_i\psi\,
 \overline{\psi}T_{\mu\nu\, i}\,\psi,\label{lag}
\end{eqnarray}
where the first term describes the pseudovector coupling
and the second one the tensor coupling,
\[
 \mitg^\mu_i=\gamma_5\gamma^\mu\tau_i \,,\qquad
T^{\mu\nu}_i=\sigma^{\mu\nu}\tau_i.
\]
We have also defined the isospin operators for convenience as
\[
\tau_\pm=
\frac{\tau_x \pm i\tau_y}{\sqrt{2}}\, \ \ ,\qquad
\tau_0=\tau_z.
\]
The coupling constants $g_a$ and $g_t$ will be related
later to the LM parameter $g'$ in non-relativistic models.

The RPA correlation function for the external field, $\mitg_A$
and $\mitg_B$, is given by the mean field correlation
function ${\mit\Pi}$ as\cite{ksg1,ksg2,ks1}
\begin{eqnarray}
{\mit\Pi}_{\rm RPA}(\mitg_A,\mitg_B\,)
& = &{\mit\Pi}(\mitg_A,\mitg_B\,) \nonumber \\
& &
+\,\chi_\alpha\,
{\mit\Pi}(\mitg_A,\mitg^\alpha_i\,)\,
{\mit\Pi}_{\rm RPA}(\mitg_{\alpha i},\mitg_B\,),\ \ \ \ \label{pirpa}
\end{eqnarray}
where we have used the notations,
\begin{eqnarray}
\mitg^\alpha_i
=\left\{
\begin{array}{l}
\gamma_5\gamma^\mu\tau_i,  \\
\noalign{\vskip4pt}
\sigma^{\mu\nu}\tau_i\ \ (\ \mu>\nu\ ), 
\end{array}
\right.
\quad
\chi_\alpha=\left\{
\begin{array}{ll}
\displaystyle{\frac{g_a}{(2\pi)^3}}\,,&(\alpha=a)\\
\noalign{\vskip4pt} 
\displaystyle{\frac{g_t}{(2\pi)^3}}\,,&(\alpha=t).
\end{array}
\right. \nonumber
\end{eqnarray} 
When the external fields are written  as
 $\mitg_A=\gamma_{\rm a}\tau_-$\,, and
$\mitg_B=\gamma_{\rm b}\tau_+$, $\gamma_{\rm a}$
and $\gamma_{\rm b}$ being some $4\times 4$ matrixes,
Eq.(\ref{pirpa}) becomes to be 
%\begin{widetext}
\begin{eqnarray}
& &
{\mit\Pi}_{\rm RPA}(\gamma_{\rm a}\tau_-,\gamma_{\rm b}\tau_+)
 \nonumber  \\
\noalign{\vskip2pt} 
&=&
{\mit\Pi}(\gamma_{\rm a}\tau_-,\gamma_{\rm b}\tau_+) \nonumber  \\
\noalign{\vskip2pt}  
& &
+\,
\chi_\alpha
{\mit\Pi}(\gamma_{\rm a}\tau_-,\mitg^\alpha_+)
{\mit\Pi}_{\rm RPA}(\mitg_{\alpha -},\gamma_{\rm b}\tau_+).
\label{pirpa_1}
\end{eqnarray}
In the same way, the last term of the above equation is
described as,
\begin{eqnarray}
& & 
{\mit\Pi}_{\rm RPA}(\mitg_{\alpha -},\gamma_{\rm b}\tau_+) \nonumber  \\
\noalign{\vskip2pt}
&=&
{\mit\Pi}(\mitg_{\alpha -},\gamma_{\rm b}\tau_+) \nonumber  \\
\noalign{\vskip2pt}
& &
+\,\chi_\beta
{\mit\Pi}(\mitg_{\alpha -},\mitg^\beta_+\,)\,
{\mit\Pi}_{\rm RPA}(\mitg_{\beta -},\gamma_{\rm b}\tau_+).
\end{eqnarray}
This can be rewritten as,
\begin{eqnarray}
{\mit\Pi}_{\rm RPA}(\mitg_{\alpha -},\gamma_{\rm b}\tau_+\,)
=\left(U^{-1}\right)_{\alpha\beta}
{\mit\Pi}(\mitg_{\beta -},\gamma_{\rm b}\tau_+\,),\label{pirpa_2}
\end{eqnarray}
with use of the dimesic function,
\begin{eqnarray}
 U_{\alpha\beta}=\delta_{\alpha \beta}
 -\chi_\beta\,
{\mit\Pi}(\mitg_{\alpha -},\mitg^\beta_+\,).
\end{eqnarray}
From Eq.(\ref{pirpa_2}),
Eq.(\ref{pirpa_1}) is expressed in the form,
\begin{eqnarray}
& &
 {\mit\Pi}_{\rm RPA}(\gamma_{\rm a}\tau_-,\gamma_{\rm b}\tau_+) \nonumber  \\
\noalign{\vskip2pt}
&=&
{\mit\Pi}(\gamma_{\rm a}\tau_-,\gamma_{\rm b}\tau_+) \nonumber  \\
\noalign{\vskip2pt}
& &
+\,\chi_\alpha
{\mit\Pi}(\gamma_{\rm a}\tau_-,\mitg^\alpha_+)
\left(U^{-1}\right)_{\alpha\beta}
{\mit\Pi}(\mitg_{\beta -},\gamma_{\rm b}\tau_+).
\end{eqnarray}
%\end{widetext}
Thus, ${\mit\Pi}_{\rm RPA}$ can be described in terms of the mean
field correlation function ${\mit\Pi}$. 
 
When we calculate the mean field correlation function,
we neglect the divergent terms and keep all the nuclear density
dependent terms including the Pauli blocking ones,
as most of the previous authors did\cite{chin,ks2}.
We call this approximation no free term
approximation (NFA)\cite{ks1}.
As discussed in details in Ref.\cite{ks1},
the Pauli blocking terms are
necessary for keeping at least the IFF sum rule and
the current conservation\cite{ks1,ks3}.
The no-sea approximation (NSA)\cite{dawson} is equivalent to NFA 
in the description of the low lying states where
nucleon-degrees of freedom play a main role\cite{ks1}.
In the following calculations, we will use NFA, but
will come back to the problem of this approximation
at the end of Sec.\ref{strength}.

In NFA,
the mean field correlation function is given by
\begin{eqnarray}
& &{\mit\Pi}( \gamma_a\tau_- , \gamma_b\tau_+\,) \nonumber \\
\noalign{\vskip2pt} 
&=& 
\int\!\frac{d^3p}{\ep{p}}
\left(
\frac{t_{ab}(p,q)}{(p+q)^2-M^{\ast2}+i\varepsilon}
\,\thn{p} \right. \nonumber \\
& &
\phantom{
\int\!\frac{d^3p}{\ep{p}}
} 
\left. 
+\,\frac{t_{ba}(p,-q)}{(p-q)^2-M^{\ast2}+i\varepsilon}
\,\thp{p}
\right) \nonumber \\
 \noalign{\vskip2pt}
& &
+\, i\pi\!\!
\int\!\!d^3p\,\frac{
\delta(q_0+E_{\vct{p}}-E_{\vct{p}+\vct{q}} )}{\ep{p} E_{\vct{p}+\vct{q}} }
\,t_{ab}(p,q)\thn{p}\thp{p+q},\ \ \ \ \ 
\label{pimf}
\end{eqnarray}
where $q$ stands for the four-momentum transfer from the external field to
the nucleus. The notations are defined as
$p_0=E_{\vct{p}}=\sqrt{\vct{p}^2+M^{\ast 2}}$,
$M^\ast$ being the nucleon effective mass,
and $\thi{p}=\theta(\kfi -|\vct{p}|)$, $\kfn$ and $\kfp$
being the Fermi momentum of the neutrons and protons, respectively. 
The step functions express the density-dependence.
We have also used the abbreviation in the above equation,
\begin{eqnarray}
t_{ab}(p,q)=-\,
{\rm Tr}_\sigma\Bigl(
\gamma_a\left(\fslash{p}\,\,+\fslash{q}+M^\ast\right)
\gamma_b\left(\fslash{p}+M^\ast\right)
\Bigr).\label{tab}
\end{eqnarray}
In the present paper, we study the GT transition at $\vct{q}=0$.
In this case, Eq.(\ref{tab}) can be written, as shown later
explicitly, in the following form, 
\begin{eqnarray}
 t_{ab}(p,q)=f(p)+g(p)q_0.
\end{eqnarray}
Then, Eq.(\ref{pimf}), for $N\ge Z$ nuclei, can be expressed
in terms of $f(p)$ and $g(p)$ as
\begin{eqnarray}
& &{\mit\Pi}( \gamma_a\tau_- , \gamma_b\tau_+\,) \nonumber \\
\noalign{\vskip2pt}
&= &
\int\!\frac{d^3p}{2p_0^2}
\left[
\frac{f}{q_0+i\varepsilon} 
\left(\thn{p}-\thp{p} \right) \right.  \nonumber \\
& &
\left.
+\left(2p_0g - f \right)
\left(
 \frac{\thn{p}}{2p_0+q_0-i\varepsilon}
 +\frac{\thp{p}}{2p_0-q_0-i\varepsilon}
\right)
\right].  \nonumber
\end{eqnarray}
For giant GT states with $|q_0|\ll M^\ast$,
the above equation may be written as
\begin{eqnarray}
 {\mit\Pi}( \gamma_a\tau_- , \gamma_b\tau_+\,)
&=& 
 \int\!\frac{d^3p}{2p_0^2}
\left[\,
\frac{f}{q_0+i\varepsilon}
\left(\thn{p}-\thp{p} \right) \right. \nonumber \\
\noalign{\vskip2pt} 
& &
\left. 
+\,\frac{2p_0g-f}{2p_0}
\left(
\thn{p}
+\thp{p}
\right)
\right]. \ \ \  \label{pi_2}
\end{eqnarray}

As shown in the previous paper\cite{ksg2},
the GT states can be described
by taking the component, $\gamma_a=\gamma_b=\gamma_5\gamma_2$,
as the external field. 
This component has non-zero values of $t_{ab}$ in the following
correlation functions,
\begin{eqnarray}
{\mit\Pi}_{11}
 &=&{\mit\Pi}(\gamma_5\gamma_2\tau_-,\gamma_5\gamma_2\tau_+)\nonumber \\
\noalign{\vskip2pt}
&=&-\,{\mit\Pi}(\gamma_5\gamma_2\tau_-,\gamma_5\gamma^2\tau_+)\\
\noalign{\vskip4pt}
{\mit\Pi}_{12}
 &=&{\mit\Pi}(\gamma_5\gamma_2\tau_-,\sigma_{31}\tau_+)\nonumber \\
\noalign{\vskip2pt} 
&=&{\mit\Pi}(\gamma_5\gamma_2\tau_-,\sigma^{31}\tau_+)\nonumber \\
\noalign{\vskip2pt}
&=&-\,{\mit\Pi}(\sigma_{31}\tau_-,\gamma_5\gamma^2\tau_+).
\end{eqnarray}
Therefore, in writing
\begin{eqnarray}
{\mit\Pi}_{22}
 &=&{\mit\Pi}(\sigma_{31}\tau_-,\sigma_{31}\tau_+)
 ={\mit\Pi}(\sigma_{31}\tau_-,\sigma^{31}\tau_+),
\end{eqnarray}
the dimesic function is provided as
\begin{eqnarray}
 U&=&\delta_{\alpha \beta}
 -\chi_\beta\,
{\mit\Pi}(\mitg_{\alpha -},\mitg^\beta_+\,)\nonumber \\
&=&\left(
\begin{array}{cc}
1+\chi_a{\mit\Pi}_{11} & -\,\chi_t{\mit\Pi}_{12} \\
\chi_a{\mit\Pi}_{12}   & 1 -\chi_t{\mit\Pi}_{22}
\end{array}
\right).\label{dimesic}
\end{eqnarray}

Calculations of $t_{ab}$ for ${\mit\Pi}_{ij}$ are straightforward.
For ${\mit\Pi}_{11}$, we obtain
\[
 t_{ab}=-\,4\left(2M^{\ast2}+2p_y^2+E_{\vct{p}}q_0 \right).
\]
This, together with Eq.(\ref{pi_2}), gives
\begin{eqnarray}
	{\mit\Pi}_{11}&=&
	 -\,\frac{4}{q_0+i\varepsilon}\int\!d^3p\,
	 \frac{M^{\ast2}+p_y^2}{E_{\vct{p}}^2}
	 \left( \thn{p}-\thp{p} \right)\nonumber \\
 \noalign{\vskip2pt}
& &	 -\,2\int\!d^3p\,
	 \frac{\vct{p}^2-p_y^2}{E_{\vct{p}}^3}
	 \left( \thn{p}+\thp{p} \right).    
 \end{eqnarray}
In the case of ${\mit\Pi}_{12}$, $t_{ab}$ is calculated to be
\[
 t_{ab}=-\,4M^\ast\left(2E_{\vct{p}}+q_0 \right),
\]
which yields
\begin{eqnarray}
	{\mit\Pi}_{12}
	 =-\,
	 \frac{4M^\ast}{q_0+i\varepsilon}
	 \int\!d^3p\,
	 \frac{\thn{p}-\thp{p}}{E_{\vct{p}}}.
\end{eqnarray}
In the same way, ${\mit\Pi}_{22}$ is described as
 \begin{eqnarray}
	{\mit\Pi}_{22}&=&
	 -\,\frac{4}{q_0+i\varepsilon}\int\!d^3p\,
	 \frac{M^{\ast2}+2p_y^2}{E_{\vct{p}}^2}
	 \left( \thn{p}-\thp{p} \right)\nonumber \\
  & &
	 -\,2\int\!d^3p\,
	 \frac{\vct{p}^2-2p_y^2}{E_{\vct{p}}^3}
	 \left( \thn{p}+\thp{p} \right),	
\end{eqnarray}
by using
\[
 t_{ab}=-\,4\left(2M^{\ast2}+4p_y^2+E_{\vct{p}}q_0 \right).
\]

In order to express the above equations in a simpler form,
let us employ the following notations,
\begin{eqnarray}
Q_a(\kf)&=&4\int_0^{\kf}\!\!d^3p\,\frac{M^{\ast2}+p_y^2}{E_{\vct{p}}^2}\,,\\
\noalign{\vskip2pt}
Q_m(\kf)&=&4\int_0^{\kf}\!\!d^3p\,\frac{M^\ast}{E_{\vct{p}}}\,,\\
\noalign{\vskip2pt}
Q_t(\kf)&=&4\int_0^{\kf}\!\!d^3p\,\frac{M^{\ast2}+2p_y^2}{E_{\vct{p}}^2}\,,\\
\noalign{\vskip2pt}
\kappa&=&\frac23\int\!d^3p\,\frac{\vct{p}^2}{E_{\vct{p}}^3}\label{kappa}
\left( \thn{p}+\thp{p} \right)\,,
\end{eqnarray}
and, moreover,
\begin{eqnarray}
q_a&=&Q_a(\kfn)-Q_a(\kfp)\,,\label{q}\\
\noalign{\vskip2pt}
q_m&=&Q_m(\kfn)-Q_m(\kfp)\,,\label{q_m}\\
\noalign{\vskip2pt}
q_t&=&Q_t(\kfn)-Q_t(\kfp).\label{q_t}
\end{eqnarray}
Then, ${\mit\Pi}_{ij}$ is described as
\begin{eqnarray}
{\mit\Pi}_{11}&=&-\,\frac{q_a}{q_0+i\varepsilon}-2\kappa\,, \\
\noalign{\vskip2pt}
{\mit\Pi}_{12}&=&-\,\frac{q_m}{q_0+i\varepsilon}\,,\\
\noalign{\vskip2pt}
{\mit\Pi}_{22}&=&-\,\frac{q_t}{q_0+i\varepsilon}-\kappa.
\end{eqnarray}

The functions in Eq.(\ref{q})  can be expanded
in terms of $(\kfn -\kfp)$, for example, as
\[
 q_a \approx \frac{dQ(\kf)}{d\kf}(\kfn - \kfp), \quad
 \frac{dQ(\kf)}{d\kf}=16\pi\kf^2\left(1-\frac{2}{3}\vf^2\right),
\]
where $\vf$ stands for the Fermi velocity, $\kf / \ekf$, with
$\ekf = \sqrt{\kf^2+M^{\ast 2}}$.
When we use the relationship, as usual,
\[
 \kfn - \kfp \approx \frac{2}{3}\kf\frac{N-Z}{A},
\]
the function $q_a$ is written in terms of $(N-Z)$ as
\begin{eqnarray}
q_a\approx \alpha\left(1-\frac23\vf^2\right)\,,\quad
 \alpha=32\pi\kf^3\frac{N-Z}{3A}.\label{qa}
\end{eqnarray}
In the same way, the functions, $q_m$ and $q_t$,
are expressed approximately as
\begin{eqnarray}
q_m\approx \alpha\sqrt{1-\vf^2}\,,\quad
q_t\approx \alpha\left(1-\frac13\vf^2\right).\label{qt}
\end{eqnarray}
We note again that the above equations are obtained by
their expansion in terms of $(N-Z)$, but not
in terms of $\vf$.
On the other hand, if we expand
$\kappa$ in Eq.(\ref{kappa}) in terms of $\vf$, we obtain
\begin{eqnarray}
\kappa \approx \frac{8\pi}{15}\kf^2\vf^3\left(1+\frac{3}{7}\vf^2
+\cdots \right).
\end{eqnarray}
This comes from Pauli blocking terms due to
nucleon-antinucleon excitations\cite{ksg2,ks1}. Since 
$\kappa$ is of order $\vf^3$, the Pauli
blocking effect is negligible in the case
of the GT excitations\cite{ksg2,ks1}.

\section{Excitation Energy of the GT States}\label{energy}

The excitation energy of the GT state is determined by the determinant
of the dimesic function,
\begin{eqnarray}
 {\rm det}\,U=0 \label{eigen},
\end{eqnarray}
which gives $q_0=\omega_{\pm}$,
\begin{eqnarray}
\omega_\pm=\frac12\left(
 \omega_a+\omega_t
 \pm\sqrt{(\omega_a-\omega_t)^2
 -4\tilde{\chi}_a\tilde{\chi}_t q_m^2
 }
 \,\right)
 \label{fulleigen}
\end{eqnarray}
with
\begin{eqnarray}
\omega_a=
\tilde{\chi}_aq_a\,,\quad \omega_t =-\,\tilde{\chi}_tq_t
\,, \\
\noalign{\vskip4pt} 
\tilde{\chi}_a=\frac{\chi_a}{1-2\kappa \chi_a}
\,,\quad \tilde{\chi}_t=\frac{\chi_t}{1+\kappa \chi_t}.
\end{eqnarray}
It is obvious that $\omega_a$ and $\omega_t$ are
the solutions of Eq.(\ref{eigen}),
when there is no mixing between the pseudovector and tensor couplings.  
According to Eqs.(\ref{qa}) and (\ref{qt}),
they are given approximately as 
\begin{eqnarray}
\omega_a \approx
 \alpha \tilde{\chi}_a\left(1-\frac23\vf^2\right)\,,\quad
\omega_t \approx
 -\,\alpha \tilde{\chi}_t\left(1-\frac13\vf^2\right).\label{rel}
\end{eqnarray}
In non-relativistic models, it has been shown that
the LM parameter $g'$ provides us with the excitation energy of
the GT state\cite{suzuki},
\begin{eqnarray}
\omega_{\rm GT}=g'\left(\frac{f_\pi}{m_\pi}\right)^2
 \frac{4\kf^3}{3\pi^2}\frac{N-Z}{A}.\label{nonrel}
\end{eqnarray}
Comparing the above equation to Eq.(\ref{rel}),
it is seen that both pseudovector and tensor coupling can reproduce
individually
the non-relativistic result by the relationship,
\begin{eqnarray}
g_a=g'\left(\frac{f_\pi}{m_\pi}\right)^2\,,\qquad
g_t=-g'\left(\frac{f_\pi}{m_\pi}\right)^2,
\end{eqnarray}
but they produce a different relativistic correction of order
$\vf^2$ from each other. The above relationship is also
verified from the comparison of Eq.(\ref{LM}) with the one from
the non-relativistic reduction of the  space
part of the Lagrangian
Eq.(\ref{lag}), 
\begin{eqnarray}
 {\cal L}\approx
-\,\frac{g_a-g_t}{2}
\psi^\dagger\gvct{\sigma}\tau_i\psi\cdot
\psi^\dagger\gvct{\sigma}\tau_i\psi.
\end{eqnarray}

If we take both the pseudovector and tensor couplings,
$\omega_+$ in Eq.(\ref{fulleigen})
up to second order of $\vf$ is written as
\begin{eqnarray}
\omega_+\approx
\alpha
\left(
\chi_a-\chi_t
-\frac{2\chi_a-\chi_t}{3}\vf^2
\right),\label{gt+}
\end{eqnarray}
while $\omega_-$ is of fourth order,
\begin{eqnarray}
\omega_-\approx -\,\alpha
\frac{2\chi_a\chi_t}
{9\left(\chi_a-\chi_t\right)}
\vf^4.
\end{eqnarray}
As mentioned in Sec.\ref{intro}, the non-relativistic LM parameter
works in a way that a half of $g'$ is for
the $\pi$-meson exchange potential, and the rest is for
the $\rho$-meson exchange one. In this sense, it may be reasonable
to use the pseudovector and the tensor coupling with equal weight
in relativistic models.
In assuming that 
\begin{eqnarray}
 g_a=-g_t=\frac{1}{2}g'\left(\frac{f_\pi}{m_\pi}\right)^2\,
\end{eqnarray}
the above $\omega_+$ becomes
\begin{eqnarray}
\omega_+\approx \omega_{\rm GT}\left(1-\frac{1}{2}\vf^2 \right).
\end{eqnarray}

\section{The Excitation Strength of the GT States}\label{strength}

The GT strength is given by the imaginary part of the RPA correlation
function. The RPA correlation function is written,
using Eq.(\ref{pirpa_1}), as
\[
  {\mit\Pi}_{\rm RPA}(\gamma_5\gamma_2\tau_-,\gamma_5\gamma_2\tau_+)
 = \left(U^{-1}\right)_{1\alpha}
{\mit\Pi}(\mitg_{\alpha-},\gamma_5\gamma_2\tau_+\,).\nonumber 
\]
The explicit expression of the dimesic function in Eq.(\ref{dimesic})
yields
\begin{eqnarray}
 {\mit\Pi}_{\rm RPA}(\gamma_5\gamma_2\tau_-,\gamma_5\gamma_2\tau_+)
=
\frac{1}{\chi_a}
\left(1
-\frac{1-\chi_t{\mit\Pi}_{22}}{{\rm det}U}
\right).
\end{eqnarray}
The last term in the parentheses can be expressed in terms of the
eigenvalues in the preceding section,
\begin{eqnarray}
& &\frac{1-\chi_t{\mit\Pi}_{22}}{{\rm det}U} \nonumber  \\
\noalign{\vskip2pt} 
&=&
\frac{1}{1-2\kappa \chi_a}
\left(
1
+\frac{\omega_a-\omega_-}{\omega_+ - \omega_-}
\frac{\omega_+}{q_0-\omega_+ +i\varepsilon} \right. \nonumber  \\
& &
\phantom{
\frac{1}{1-2\kappa \chi_a}
1+
} 
\left.
+\,\frac{\omega_+ - \omega_a}{\omega_+ - \omega_-}
\frac{\omega_-}{q_0-\omega_- +i\varepsilon}
\right).
\end{eqnarray}
The response function for the external field,
$\gamma_5\gamma_2\tau_+$, therefore, is described as
\begin{eqnarray}
& &R_a(q_0)\nonumber \\
\noalign{\vskip2pt}  
&=&\frac{3}{16\pi^2}\frac{A}{\kf^3}\nonumber
\,{\rm Im}\,{\mit\Pi}_{\rm RPA} \\
\noalign{\vskip2pt}
&=&
\frac{3}{16\pi}\frac{A}{\kf^3}
\frac{1}{\chi_a\left( 1-2\kappa \chi_a \right)}
\left(
\frac{\omega_a-\omega_-}{\omega_+ - \omega_-}
\omega_+ \delta(q_0-\omega_+)
\right. \nonumber \\
& &
\phantom{
\frac{A}{\kf^3}
\frac{1}{\chi_a\left( 1-2\kappa \chi_a \right)}
} 
\left.
+\,
\frac{\omega_+-\omega_a}{\omega_+
- \omega_-} \omega_- \delta(q_0-\omega_-)
\right).\ \ \ \ 
\end{eqnarray}
The above equation provides us with the GT strengths of the
two states with the excitation energy, $\omega_+$ and $\omega_-$,
respectively,
\begin{eqnarray}
S_+&=&\frac{3}{16\pi}\frac{A}{\kf^3}
\frac{1}{\chi_a\left( 1-2\kappa \chi_a \right)}
\frac{\omega_a-\omega_-}{\omega_+ - \omega_-}\,\omega_+\,,\label{sp}\\
\noalign{\vskip4pt}
S_-&=&\frac{3}{16\pi}\frac{A}{\kf^3}
\frac{1}{\chi_a\left( 1-2\kappa \chi_a \right)}
\frac{\omega_+-\omega_a}{\omega_+ - \omega_-}\,\omega_- \,.
\label{sm}
\end{eqnarray}
If we take into account order up to $\vf^3$, they become
\begin{eqnarray}
S_+&\approx&
\frac{3}{16\pi}\frac{A}{\kf^3}
\frac{\omega_a}{\chi_a\left( 1-2\kappa \chi_a \right)}\nonumber \\
\noalign{\vskip2pt}
&=&\frac{1}{\left( 1-2\kappa \chi_a \right)^2}
\left(1-\frac23\vf^2\right)2\left(N-Z\right)\,,\label{gt}\\
\noalign{\vskip4pt}
S_-&\approx& 0.
\end{eqnarray}
Thus, the strength of the GT state does not depend on the tensor
coupling, up to $\vf^3$, although the excitation energy does,
as shown in the preceding section.
The sum of the two strengths in Eqs.(\ref{sp}) and (\ref{sm}) is
given, without expansion in terms of $\vf$, as
\begin{eqnarray}
S_+ + S_- &=& \int\!dq_0\,R_a(q_0)  \nonumber \\
&=&\frac{3}{16\pi}\frac{A}{\kf^3}
\frac{\omega_a}{\chi_a\left( 1-2\kappa \chi_a \right)}\nonumber \\
&=&\frac{1}{\left( 1-2\kappa \chi_a \right)^2}
\left(1-\frac23\vf^2\right)2\left(N-Z\right),\quad \label{psres}
\end{eqnarray}
which is independent of the tensor coupling.
If we discuss the GT strength in the nucleon space only, then
the sum is independent of $g_a$ also,
because of $\kappa=0$,
and is quenched by the
factor $(1-2\vf^2/3)$, compared with IFF sum rule value $2(N-Z)$
in the present definition\cite{ksg2, ks1}. The last equation is
what we obtained in our previous paper\cite{ksg1,ksg2,ks1}. 

It may be worthwhile noting the response function for the external
field, $\sigma_{31}\tau_+$. The RPA correlation function in this
case can be written as
\begin{eqnarray}
& &{\mit\Pi}_{\rm RPA}(\sigma_{31}\tau_-,\sigma_{31}\tau_+)  \nonumber \\
&=&
-\,\frac{\kappa}{1+\kappa\chi_t}  \nonumber \\
& &
+\,\frac{1}{\chi_t\left(1+\kappa\chi_t\right)}
\left(
\frac{\omega_t-\omega_-}{\omega_+ - \omega_-}
\frac{\omega_+}{q_0-\omega_+ +i\varepsilon} \right. \nonumber \\
& &
\left.
\phantom{
+\frac{1}{\left(1+\kappa\chi_t\right)}
} 
+\frac{\omega_+ - \omega_t}{\omega_+ - \omega_-}
\frac{\omega_-}{q_0-\omega_- +i\varepsilon}
\right).
\end{eqnarray}
This gives the response function,
\begin{eqnarray}
& &
R_t(q_0) \nonumber \\
\noalign{\vskip2pt}
&=&\frac{3}{16\pi^2}\frac{A}{\kf^3}
\,{\rm Im}\,{\mit\Pi}_{\rm RPA} \nonumber \\
\noalign{\vskip2pt}
&=&
-\,\frac{3}{16\pi}\frac{A}{\kf^3}
\frac{1}{\chi_t\left( 1+\kappa \chi_t \right)}
\left(
\frac{\omega_t-\omega_-}
{\omega_+ - \omega_-}\,\omega_+\,\delta(q_0-\omega_+) \right. \nonumber \\
& &
\left. 
\phantom{
\frac{A}{\kf^3}
\frac{1}{\chi_t\left( 1+\kappa \chi_t \right)}
}
+\,\frac{\omega_+-\omega_t}
{\omega_+ - \omega_-}\,\omega_-\,\delta(q_0-\omega_-)
\right).
\end{eqnarray}
The sum of the strengths expanded in terms of $(N-Z)$,
therefore, is obtained as 
\begin{eqnarray}
\int\!dq_0\,R_t(q_0)
&=&
-\,\frac{3}{16\pi}\frac{A}{\kf^3}
\frac{\omega_t}{\chi_t\left( 1+\kappa \chi_t \right)} \nonumber \\
\noalign{\vskip2pt} 
&=&\frac{1}{\left( 1+\kappa \chi_t \right)^2}
\left(1-\frac13\vf^2\right)2\left(N-Z\right),\ \ \ \ \label{tenres}
\end{eqnarray}
which is independent of the pseudovector coupling.
As expected, the relativistic correction in Eq.(\ref{tenres})
is different from that in Eq.(\ref{psres}).

Finally effects of the Dirac sea in the relativistic model
should be mentioned in more detail.
In the above calculations, effects of the Dirac sea are included
in $\kappa$ in Eq.(\ref{kappa}), which comes from the Pauli blocking
terms in NFA.
Since its value is positive,
the GT strength in Eqs.(\ref{gt}) and (\ref{psres}) is a little
increased owing to the coupling of
the particle-hole states with the nucleon-antinucleon ones. This
increasing is, however, due to a poor approximation of NFA where the
divergent term is simply neglected. The no-sea approximation, which has
been extensively used so for, is essentially the same as NFA, and
provides us with Eqs.(\ref{gt}) and (\ref{psres}),
as shown in Ref.\cite{ks1}.
If we keep the divergent term in the RPA correlation function,
$\kappa$ in Eq.(\ref{kappa}) is replaced by $\kappa'\,$\cite{ks1},
\begin{eqnarray}
\kappa'&=&\frac23\int\!d^3p\,\frac{\vct{p}^2}{E_{\vct{p}}^3}
\left( \thn{p}+\thp{p}-2 \right).\label{kkappa}
\end{eqnarray}
Thus, the divergent term has an opposite sign to that of the
density-dependent part.
This fact implies that if we treat the divergent terms properly,
the GT strength may be quenched more than in Eqs.(\ref{gt})
and (\ref{psres}). In this sense, the factor $(1-2\vf^2/3)$
yields the minimum quenching of the present relativistic model.
We note that in nuclear matter
the quenching due to the antinucleon-degrees of freedom
is caused only through the pseudovector coupling.
In finite nuclei,
where the momentum is not a good quantum number, there 
may be a small contribution from the tensor coupling. 
It is important future work to investigate
how much the quenching is increased  
by the coupling of the particle-hole states
with the nucleon-antinucleon states.
In that case, it may be also required to take into account
the energy-dependence of the LM parameter which has not been
well studied yet.

\section{Conclusions}
In the previous papers\cite{ksg1,ksg2,ks1},
we investigated the excitation energy and
strength of the giant Gamow-Teller (GT) states in the
relativistic model by introducing the Landau-Migdal (LM) parameter
$g_a$ with the pseudovector coupling. The pseudovector coupling
is chosen so as to reproduce the non-relativistic LM parameter
$g'$ in the non-relativistic limit. The main conclusions were
that the relativistic correction to the excitation energy is 14\%
and that the GT strength in the nucleon sector is quenched by 12\%
in nuclear matter,
compared with the Ikeda-Fujii-Fujita (IFF) sum rule value.
The quenching factor is given by $(1-2\vf^2/3)$, $\vf$
being the Fermi momentum.
In finite nuclei, the quenched amount has been estimated
to be about 6\% in Ref.\cite{ksg2,ma}.
The reduction of the quenching is due to the larger  effective mass
near the nuclear surface than in nuclear matter.
Among the above results, the quenching of the strength 
is especially important. 
This prediction is a rare example to distinguish
the relativistic model from conventional non-relativistic
models, and is related to an important observation
of the recent experiment\cite{sakai}.

Recently, Tokyo group has observed 10\% quenching of the IFF
sum rule value in $^{90}$Nb
through charge-exchange reaction\cite{sakai}.   
Since nuclear models assuming the nucleus to be composed
of nucleons satisfy the model-independent sum rule, the observed
quenching implies that additional degrees of freedom play a role
to reduce the GT strength. So far, one candidate has been proposed,
which is the $\Delta$-degrees of freedom\cite{ss,arima}.
Particle-hole states are assumed to couple with $\Delta$-hole states
through the Landau-Migdal (LM) parameter $g'_{{\rm N}\Delta}$.
Since the coupling force is expected to be
repulsive, a part of the strength in the
nucleon space is taken by the highly excited states.
Of course, it depends on the value of
$g'_{{\rm N}\Delta}$ how much strength is taken out
from the low excitation energy region.  

The relativistic model predicted another source
of the quenching.
If the 6\% quenching is due to the relativistic effect,
as estimated in the previous papers\cite{ksg2,ma},
then it is about 4\% what is left for the coupling
with the $\Delta$-hole states.
In this case, the value of $g'_{{\rm N}\Delta}$
becomes less than 0.2, which is much smaller
than the one estimated before\cite{ss,arima,ksg3}.
The small value changes our previous understanding of
the spin-dependent structure of nuclei.
For example, the critical density of the
pion condensation becomes less than two times of the
normal density\cite{sst}.
Thus, it was required to investigate model-dependence of
the previous result\cite{ksg1,ksg2} in the relativistic model.

In the present paper, we have explored  whether or not the LM
parameter with the tensor coupling changes the previous conclusions.
The tensor coupling is another candidate which can
reproduce the non-relativistic result of the GT states.
We have described the GT states in the two cases. One is
to use the tensor coupling instead of the pseudovector coupling.
The other is to take into account 
both pseudovector and tensor couplings.

If we use the only tensor coupling,
the relativistic effect on the excitation energy is
reduced by a half, as shown in Eq.(\ref{rel}).
Eq.(\ref{gt}), on the other hand, shows that 
the GT strength is almost independent of the
tensor coupling,  and is quenched by the same amount
as in the case of the pseudovector coupling.

When we take into account both the pseudovector and tensor
coupling, the relativistic effect on the excitation energy
of the GT state depends on their ratio, as in Eq.(\ref{gt+}),
while the quenching of the GT strength remains in the same way
as in other cases. This fact is shown
by the factor $(1-2\vf^2/3)$ in Eq.(\ref{psres}).

In conclusion, the pseudovector and tensor coupling play
a role to determine the excitation energy of the GT state
in a similar way,
but do not change the previous conclusion\cite{ksg1,ksg2}
that the GT strength in the relativistic model is quenched
by 12\% in nuclear matter and by 6\% in finite nuclei,
compared with the IFF sum rule value.
The quenched amount may be increased by the coupling
of the particle-hole states with nucleon-antinucleon states
through the pseudovector coupling,
as discussed at the end of the preceding section.
The present result is consistent with the recent experiment\cite{sakai},
but in future work, other spin-dependent structure of nuclei, 
like the pion condensation
and response functions of the charge-exchange reaction\cite{ichimura},
should be understood consistently.
It is also important to study the present relativistic effects
on the $\beta$ decay for its precise discussions in neutrino physics.

\acknowledgments

The authors would like to thank Dr. N. Van Giai and
Professor Z. Y. Ma for useful discussions.


\begin{thebibliography}{99}
\bibitem{brown} S. -O. B$\ddot{{\rm a}}$ckman, G. E. Brown,
	and J. A. Niskanen,
	Phys. Rep. {\bf 124}, 1 (1985).
\bibitem{ss}T. Suzuki and H. Sakai, Phys. Lett. B {\bf 455}, 25 (1999). 
\bibitem{shimizu}A. Arima, T. Cheon, K. Shimizu, H. Hyuga, and
	T. Suzuki, Phys. Lett. B {\bf 122}, 126 (1983).
\bibitem{arima}A. Arima, W.Bentz, T. Suzuki, and T. Suzuki,
	Phys. Lett. B {\bf 499}, 104 (2001). 
\bibitem{serot}B. D. Serot and J. D. Walecka, Adv. Nucl. Phys.
	{\bf 16}, 1 (1986).
\bibitem{ring}P. Ring, Prog. Part. Nucl. Phys. {\bf 37}, 193 (1996).
\bibitem{conti}C. De Conti, A. P. Gale$\tilde{{\rm a}}$o,
	and F. Krmpoti$\acute{{\rm c}}$, Phys. Lett. B
	{\bf 444}, 14 (1998); {\bf 494}, 46 (2000).
\bibitem{ksg1}H. Kurasawa, T. Suzuki, and N. Van Giai,
	Phys. Rev. Lett. {\bf 91}, 062501 (2003). 
\bibitem{ksg2}H. Kurasawa, T. Suzuki, and N. Van Giai,
	Phys. Rev. C {\bf 68}, 064311 (2003), Nucl. Phys. {\bf A731},
	114 (2004). 
\bibitem{ks1}H. Kurasawa, and T. Suzuki,
	Phys. Rev. C {\bf 69}, 014306 (2004). 
\bibitem{ikeda}K. Ikeda, S. Fujii, and J. I. Fujita, Phys. Lett. {\bf 3},
	271 (1963). 
\bibitem{sakai}H. Sakai and K. Yoko, Nucl. Phys. {\bf A731}, 94 (2004).
\bibitem{chin}S. A. Chin, Ann. Phys. (N.Y.) {\bf 108}, 301 (1977).
\bibitem{ks2}H. Kurasawa and T. Suzuki, Nucl. Phys. {\bf A445}, 685 (1985).
\bibitem{ks3}H. Kurasawa and T. Suzuki, Phys. Lett. B {\bf 474}, 262 (2000).
\bibitem{dawson}J. F. Dawson and R. J. Furnstahl, Phys. Rev. C {\bf 42},
	2009 (1990).
\bibitem{suzuki}T. Suzuki, Nucl. Phys. {\bf A379}, 110 (1982).
\bibitem{ma}Z. Y. Ma, B. Q. Chen, N. Van Giai, and T. Suzuki, to appear in
	Euro. J. Phys. A (2004). 
\bibitem{ksg3}H. Kurasawa, T. Suzuki, and N. Van Giai, Nucl. Phys. {\bf
	A731}, 114 (2004).
\bibitem{sst}T. Suzuki, H. Sakai, and T. Tatsumi, in proceedings of the
	RCNP International Symposium on Nuclear Response and Medium
	Effects, edited by T. Noro, H. Sakaguchi, H. Sakai, and
	T. Wakasa (Universal Academic Press, Tokyo, 1999), p.77.
\bibitem{ichimura}K. Kawahigashi, K. Nishida, A. Itabashi, and
	M. Ichimura, Phys. Rev. C {\bf 63}, 044609 (2001).	
\end{thebibliography}
\end{document}